\documentclass[pdflatex,sn-mathphys]{sn-jnl}
\usepackage[mathlines]{lineno} 

\let\oldequation\equation
\let\oldendequation\endequation
\renewenvironment{equation}
  {\linenomathNonumbers\oldequation}
  {\oldendequation\endlinenomath}

\let\oldalign\align
\let\oldendalign\endalign
\renewenvironment{align}
  {\linenomathNonumbers\oldalign}
  {\oldendalign\endlinenomath}

\def\track#1{{\color{black}#1}}

\begin{document}
\title[Second-order effects of mutation in continuous indirect reciprocity]{Second-order effects of mutation in a continuous model of indirect reciprocity}

\author[1]{\fnm{Youngsuk} \sur{Mun}}

\author*[2]{\fnm{Seung~Ki} \sur{Baek}}\email{seungki@pknu.ac.kr}

\affil[1]{\orgdiv{Department of Physics}, \orgname{Pukyong National University}, \orgaddress{\street{Yongso-ro 45}, \city{Busan}, \postcode{48513}, \country{Korea}}}

\affil*[2]{\orgdiv{Department of Scientific Computing}, \orgname{Pukyong National University}, \orgaddress{\street{Yongso-ro 45}, \city{Busan}, \postcode{48513}, \country{Korea}}}

\abstract{We have developed a continuous model of indirect reciprocity and thereby investigated effects of mutation in assessment rules. Within this continuous framework, the difference between the resident and mutant norms is treated as a small parameter for perturbative expansion. Unfortunately, the linear-order expansion leads to singularity when applied to the leading eight, the cooperative norms that resist invasion of another norm having a different behavioral rule. For this reason, this study aims at a second-order analysis for the effects of mutation when the resident norm is one of the leading eight. We approximately solve a set of coupled nonlinear equations using Newton's method, and the solution is compared with Monte Carlo calculations. The solution indicates how the characteristics of a social norm can shape the response to its close variants appearing through mutation. Specifically, it shows that the resident norm should allow one to refuse to cooperate toward the ill-reputed, while regarding cooperation between two ill-reputed players as good, so as to reduce the impact of mutation.
This study enhances our analytic understanding on the organizing principles of successful social norms.}


\maketitle


\section{Introduction}\label{sec1}

Reciprocity can work under the gaze of others~\cite{alexander1987biology}.
The intuitive idea that we tend to help someone who has helped others has been formulated as the theory of indirect reciprocity~\cite{nowak2006evolutionary}. The first mathematical analysis was carried out by considering a conditionally cooperative norm called Image Scoring (IS)~\cite{nowak1998evolution}, which assigns good reputation to those who have cooperated and prescribes cooperation to those with good reputation. However, as long as it refers only to the co-player's cooperative \emph{action}, regardless of whom he or she has met before, it is hard to justify why one has to cooperate exclusively toward those with good reputation~\cite{clark2020indirect}. Conditional cooperation should be based on the co-player's \emph{reputation}~\cite{leimar2001evolution,panchanathan2003tale}, and one has to be allowed to refuse to cooperate if the co-player is ill-reputed. In this regard, `Standing'~\cite{sugden1986economics} has been proposed as an alternative. Experimental studies show that people tend to regard defection as bad even when it is toward someone who keeps defecting, as IS does~\cite{milinski2001cooperation}. At the same time, their assessment is significantly affected when the context of the co-player's defection is provided, consistently with Standing~\cite{bolton2005cooperation,swakman2016reputation}. In addition, such a Standing-like rule is also found in historical records on medieval trades~\cite{greif1989reputation}. A theoretical answer between IS and Standing was given by the discovery of the leading eight~\cite{ohtsuki2004should,ohtsuki2006leading} (Table~\ref{tab:eight}): They are evolutionarily stable norms that resist invasion of mutants that have different behavioral rules, and all of them assign good reputation to defection against a defector, in accordance with Standing.
To sum up, how to make a reputation system robust against behavioral mutation is now well understood.

By contrast, when multiple norms with different assessment rules are competing, such a polymorphic population is largely unexplored, although several case studies have been conducted~\cite{uchida2010effect,uchida2013effect,hilbe2018indirect}. An enumerative approach, similar to the discovery of the leading eight, shows that only unconditional defection is evolutionarily stable if error occurs~\cite{perret2021evolution}.
We have pointed out that the problem of private reputation becomes more meaningful by considering continuous reputation between good and bad, as well as a continuous spectrum of action between cooperation and defection~\cite{lee2021local,lee2022second}.
We have at least three reasons to introduce such continuity. First, empirically, a dichotomy is not enough in reporting an assessment~\cite{alwin1997feeling}. Second, in an operational sense, error introduces probabilistic mixing. For example, if one attempts to implement an action but fails with probability of $10\%$, we could say that the actual effect amounts to $90\%$ on average. The third reason is methodological: it allows us to use a small parameter with an absolute value less than one, so as to develop a perturbation theory.

\track{This work is based on a continuum formulation that we have thus developed previously~\cite{lee2021local,lee2022second}, and our purpose is to advance} a second-order theory because,
as will be explained below, the perturbation theory up to the linear order leads to singular behavior when applied to continuous versions of the leading eight.
Although the linear-order theory covers a wide range of possible norms, its failure at the leading eight deserves attention, and we wish to show how to make progress in this direction.
Our preliminary calculation has shown that numerical solutions of this second-order nonlinear system indeed give reasonable results~\cite{lee2021local}. 
From a mathematical point of view, however, a fully analytic solution is unavailable,
and the main goal of this work is to obtain an approximate solution in a closed form. Specifically, we will employ Newton's method~\cite{newman2013computational}, which has already been applied to similar problems~\cite{baek2017duality,you2017chaos}.
The resulting expression will be compared with Monte Carlo (MC) results, and we will also discuss which factors are relevant to the deviation from a cooperative initial state.

\begin{table}[h]
\begin{center}
\begin{minipage}{\textwidth}
\caption{Leading eight and IS}
\label{tab:eight}
\begin{tabular}{@{}ccccccccccccc@{}}
\toprule                                                          
& $\alpha_{1C1}$ & $\alpha_{1D1}$ & $\alpha_{1C0}$ & $\alpha_{1D0}$ &
 $\alpha_{0C1}$ & $\alpha_{0D1}$ & $\alpha_{0C0}$ & $\alpha_{0D0}$ &
 $\beta_{11}$ & $\beta_{10}$ & $\beta_{01}$ & $\beta_{00}$\\\midrule
L1 & 1 & 0 & 1 & 1 & 1 & 0 & 1 & 0 & $C$ & $D$ & $C$ & $C$\\
L2 & 1 & 0 & 0 & 1 & 1 & 0 & 1 & 0 & $C$ & $D$ & $C$ & $C$\\
L3 & 1 & 0 & 1 & 1 & 1 & 0 & 1 & 1 & $C$ & $D$ & $C$ & $D$\\
L4 & 1 & 0 & 1 & 1 & 1 & 0 & 0 & 1 & $C$ & $D$ & $C$ & $D$\\
L5 & 1 & 0 & 0 & 1 & 1 & 0 & 1 & 1 & $C$ & $D$ & $C$ & $D$\\
L6 & 1 & 0 & 0 & 1 & 1 & 0 & 0 & 1 & $C$ & $D$ & $C$ & $D$\\
L7 & 1 & 0 & 1 & 1 & 1 & 0 & 0 & 0 & $C$ & $D$ & $C$ & $D$\\
L8 & 1 & 0 & 0 & 1 & 1 & 0 & 0 & 0 & $C$ & $D$ & $C$ & $D$\\
IS & 1 & 0 & 1 & 0 & 1 & 0 & 1 & 0 & $C$ & $D$ & $C$ & $D$\\
\botrule
\end{tabular}
\begin{tablenotes}{Cooperation and defection are denoted as $C$ and $D$,
respectively, and a player's reputation is either good ($1$) or bad ($0$).
By $\alpha_{uXv}$, we mean the reputation assigned to a player who
had reputation $u$ and did $X \in \{C, D\}$ to another player with reputation $v$.
The behavioral rule $\beta_{uv}$ prescribes an action between $C$ and $D$
when the focal player and the co-player have reputations $u$ and $v$, respectively}
\end{tablenotes}
\end{minipage}
\end{center}
\end{table}

\section{Methods}\label{sec11}

\track{
Our formulation is based on the continuous model of indirect reciprocity, according to which reputation and cooperation are both treated as continuous variables~\cite{lee2021local,lee2022second}. The contribution of this work is to present an approximate solution to the set of nonlinear equations resulting from the formulation when expanded to the second order of perturbation from the initial cooperative condition.
In this section, we will recap our formulation for the sake of completeness, and one could skip this section if already well aware of the continuum formulation developed in our previous works~\cite{lee2021local,lee2022second}.}

\subsection{Donation game}

\track{
The Prisoner's Dilemma (PD) game is described by four elementary payoffs as follows:
\begin{equation}
    \left(
    \begin{array}{c|cc}
    & C & D\\\hline
    C & R & S\\
    D & T & P
    \end{array}
    \right),
\label{eq:payoff}
\end{equation}
where $2R>T+S$ and $T>R>P>S$. We have shown only the column player's payoffs because this is a symmetric game. After payoff normalization, the PD game is parametrized by two dilemma strengths, $Dg' \equiv (T - R) / (R - P)$ and $Dr' \equiv (P - S) / (R - P)$~\cite{wang2015universal,ito2018scaling,tanimoto2021sociophysics}.
We consider the donation game, a special form of the PD for which $Dr' = Dg' = c/(b-c)$, where $b$ and $c$ mean the benefit and the cost of cooperation, respectively. We believe that our theoretical framework can be extended to the general PD game as well by assigning $Dg'$ and $Dr'$ separately.
}

\track{
In the conventional setting, a donor can choose to either cooperate or defect toward a recipient. If the donor cooperates, his or her payoff decreases by $c$, while the recipient's payoff increases by $b$. If the donor defects, nothing happens to their payoffs. When $b>c>0$, this game constitutes the PD game, in which the Nash equilibrium is defection.
Our continuum formulation generalizes the game so that the donor, say, player $i$, chooses the degree of cooperation $\beta_i$ between zero and one: then, the donor's payoff decreases by $c\beta_i$, and the recipient's payoff increases by $b\beta_i$. Player $i$ determines the degree of cooperation by referring to the recipient's reputation, so we discuss the dynamics of reputation below.
}

\subsection{Dynamics of reputation}
\label{sec:dynamics}

Imagine a large population of size $N \gg 1$. Every time step, we randomly choose two players for the donation game, one as a donor and the other as a recipient. Let us denote them by $i$ and $j$, respectively.
Let $m_{ij}^t$ be player $i$'s assessment of player $j$ at time $t$. \track{This variable takes a value between zero and one: $m_{ij}^t=1$ ($0$) means that $i$ regards $j$ as perfectly good (bad), but it can also be $m_{ij}^t=0.8$ or $0.3$, for example.} From $i$'s
point of view, how much to
cooperate toward $j$ is a function of $m_{ij}^t$ and self-assessment
$m_{ii}^t$, so let us write the amount of $i$'s cooperation toward $j$ as $\beta_i
\left( m_{ii}^t, m_{ij}^t \right)$. Observer $k$ updates the assessment of $i$
according to his or her assessment rule $\alpha_k$,
if the interaction between $i$ and $j$ is observed with probability $q$: If not,
the previous assessment is preserved. Observer $k$'s
assessment rule is a function of his or her assessments of $i$ and $j$
as well as $\beta_i$, the observed amount of $i$'s cooperation, so
let this function be written as $\alpha_k
\left( m_{ki}^t, \beta_i, m_{kj}^t \right)$.
\track{Both $\alpha_k(x,y,z)$ and $\beta_i(x,y)$ are continuous functions, taking continuous variables between zero and one as input and having the unit interval as their ranges.}
The average dynamics of $m_{ki}^t$
can thus be summarized as follows:
\begin{equation}
m_{ki}^{t+1} = (1-q) m_{ki}^t + \frac{q}{N-1} \sum_{j \neq i} \alpha_k \left[
m_{ki}^t, \beta_i \left(m_{ii}^t, m_{ij}^t \right), m_{kj}^t \right],
\end{equation}
where we have taken average over $j$ because the recipient is chosen randomly
from the population of size $N$ at each interaction.
If we take continuous-time approximation, we arrive at the following differential equation:
\begin{equation}
\frac{{\text{d}}}{{\text{d}}t} m_{ki} = -qm_{ki} + \frac{q}{N-1} \sum_{j \neq i}
\alpha_k \left[ m_{ki}, \beta_i (m_{ii}, m_{ij}), m_{kj} \right].
\label{eq:dynamics}
\end{equation}
In this formulation, the observation probability $q$ only rescales the overall time scale, so we may choose $q=1$ for the sake of convenience.

\subsection{Assessment error}
\label{sec:error}

Before considering mutation, we wish to discuss how error affects our analysis.
Although assessment error does not explicitly appear in the above dynamics, we assume that it occurs with an infinitesimal rate.
Let us consider a homogeneous population in which everyone obeys a single norm so that $\alpha_k = \alpha$ and $\beta_i = \beta$. 
We assume that this resident norm has a fixed point at which everyone
cooperates and has good reputation, meaning that
\begin{equation}
\alpha(1,1,1) = \beta(1,1) = 1.
\label{eq:fixed}
\end{equation}
The system initially starts from this cooperative fixed point, where $m_{ij}=1$ for any $i$ and $j$, but the presence of assessment error introduces small deviation $\epsilon_{ij} = 1-m_{ij} \ll 1$. By expanding Eq.~\eqref{eq:dynamics} to the linear order of $\epsilon_{ij}$'s, we obtain an $N^2$-dimensional linear-algebraic system 
\begin{eqnarray}
\frac{{\text{d}}}{{\text{d}}t} \epsilon_{ki} \approx
-q(1-A_x) \epsilon_{ki} + q A_y B_x \epsilon_{ii} +
\frac{q}{N-1}
\sum_{j \neq i} [A_y B_y \epsilon_{ij} + A_z \epsilon_{kj}],
\label{eq:n2dim}
\end{eqnarray}
where
\begin{subequations}
\begin{align}
A_x &\equiv \left.\partial_x \alpha(x,y,z) \right\vert_{(1,1,1)}\\
A_y &\equiv \left.\partial_y \alpha(x,y,z) \right\vert_{(1,1,1)}\\
A_z &\equiv \left.\partial_z \alpha(x,y,z) \right\vert_{(1,1,1)}\\
B_x &\equiv \left.\partial_x \beta(x,y) \right\vert_{(1,1)}\\
B_y &\equiv \left.\partial_y \beta(x,y) \right\vert_{(1,1)}.
\end{align}
\label{eq:derivatives}
\end{subequations}

\begin{table}[h]
\begin{center}
\begin{minipage}{\textwidth}
    \caption{Continuous versions of the leading eight and IS}
\label{tab:cont}
\begin{tabular}{@{}lcc@{}}
\toprule                                                          
        Norm & $\alpha(x,y,z)$ & $\beta(x,y)$ \\\midrule
        L1 & $x+y-xy-xz+xyz$ & $-x+xy+1$\\
        L2 (Consistent Standing) & $x+y-2xy-xz+2xyz$ & $-x+xy+1$\\
        L3 (Simple Standing) & $yz - z + 1$ & $y$\\
        L4 & $-y-z+xy+2yz-xyz+1$ & $y$\\
        L5 & $-z-xy+yz+xyz+1$ & $y$\\
        L6 (Stern Judging) & $-y-z+2yz+1$ & $y$\\
        L7 (Staying) & $x-xz+yz$ & $y$\\
        L8 (Judging) & $x-xy-xz+yz+xyz$ & $y$\\
        IS (Image Scoring) & $y$ & $y$\\
\botrule
\end{tabular}
\begin{tablenotes} {The assessment rule $\alpha(x,y,z)$ is obtained from
    the tri-linear interpolation to $\alpha_{xyz}$'s in Table~\ref{tab:eight},
    with mapping $C$ and $D$ to $1$ and $0$, respectively.
    Similarly,
    the behavioral rule $\beta(x,y)$ is obtained from the bi-linear interpolation of $\beta_{xy}$'s in Table~\ref{tab:eight} \track{(see Appendix~\ref{app:interpolation} for the details of the interpolation method)}}
\end{tablenotes}
\end{minipage}
\end{center}
\end{table}

\begin{table}[h]
\begin{center}
\begin{minipage}{\textwidth}
    \caption{First-order derivatives of $\alpha(x,y,z)$ and $\beta(x,y)$ at $(x,y,z)=(1,1,1)$} 
    \label{tab:first}
\begin{tabular}{@{}cccccc@{}}
\toprule                                                          
        Norm & $A_x$ & $A_y$ & $A_z$ & $B_x$ & $B_y$ \\\midrule
        L1 & 0 & 1 & 0 & 0 & 1\\
        L2 & 0 & 1 & 1 & 0 & 1\\
        L3 & 0 & 1 & 0 & 0 & 1\\
        L4 & 0 & 1 & 0 & 0 & 1\\
        L5 & 0 & 1 & 1 & 0 & 1\\
        L6 & 0 & 1 & 1 & 0 & 1\\
        L7 & 0 & 1 & 0 & 0 & 1\\
        L8 & 0 & 1 & 1 & 0 & 1\\
        IS & 0 & 1 & 0 & 0 & 1\\
\botrule
\end{tabular}
\end{minipage}
\end{center}
\end{table}

For example, a continuous version of Image Scoring is given as
\begin{equation}
\alpha_{\text{IS}} = \beta_{\text{IS}} = y,
\label{eq:scoring}
\end{equation}
if we apply the bi- and tri-linear interpolations to the discrete version \track{as shown in Table~\ref{tab:cont} (see Appendix~\ref{app:interpolation} for the details of the interpolation method)}. The derivatives are easily computed from this expression (Table~\ref{tab:first}).
As another example, a continuous version of L3, nicknamed Simple Standing (SS), is obtained as
\begin{subequations}
\begin{align}
\alpha_{\text{SS}}(x,y,z) &= yz - z + 1\\
\beta_{\text{SS}}(x,y) &= y.
\end{align}
\label{eq:ss0}
\end{subequations}
\track{In either case, the functions $\alpha(x,y,z)$ and $\beta(x,y)$ reproduce the values in Table~\ref{tab:eight} when $x$, $y$, and $z$ are the endpoints of the input domain. For example, one can immediately see $\alpha_{\text{SS}}(1,1,1) = \alpha_{1C1}=1$, $\beta_{\text{SS}}(1,0) = \beta_{10}=0$, and so on. At the same time, the functions can now handle intermediate input values such as $\alpha_{\text{SS}}(0.3,0.6,0.5) = 0.8$.}
Note that the linear-order description for both SS and IS is given as
\begin{equation}
(A_x, A_y, A_z, B_x, B_y) = (0, 1, 0, 0, 1)
\label{eq:pure}
\end{equation}
in common (Table~\ref{tab:first}).

Equation~\eqref{eq:n2dim} can now be rewritten in a matrix form as
\begin{equation}
    \frac{{\text{d}}}{{\text{d}}t}\vec{\epsilon} = q\mathcal{M}\vec{\epsilon},
\end{equation}
where $\vec{\epsilon} \equiv \left( \epsilon_{11}, \ldots, \epsilon_{NN} \right)$. This linear system is solvable~\cite{lee2021local}, and it turns out that the growth of $\vec{\epsilon}$ is determined by the largest eigenvalue of $\mathcal{M}$, obtained as follows:
\begin{equation}
Q \equiv -1 + A_x + A_z + A_y (B_x + B_y).
\label{eq:Q}
\end{equation}
This analysis shows that the initial fixed point (Eq.~\eqref{eq:fixed}) is linearly unstable under the following four norms among the leading eight: L2, L5, L6, and L8, for which $Q = A_z >0$. \track{Note that $A_z = \alpha(1,1,1)-\alpha(1,1,0) = \alpha_{1C1}-\alpha_{1C0} = 1-\alpha_{1C0}$ according to the tri-linear interpolation so that $\alpha_{1C0}$ plays a crucial role in determining the stability.}
In addition, the corresponding eigenvector is given by $\epsilon_{11} = \ldots = \epsilon_{NN} \equiv \epsilon$, which shows that the dynamics is approximately one-dimensional~\cite{lee2022second}.

\subsection{Mutation}
\label{subsec:mutation}
Let us assume that mutation occurs in such a way that a small fraction of the population, say, $p$,
begin to use a different norm $(\alpha-\delta, \beta-\eta)$
with $\left\| \delta
\right\| \ll 1$ and $\left\| \track{\eta} \right\| \ll 1$, 
while most of the population still use a common norm defined by $\alpha_k =
\alpha$ and $\beta_i = \beta$.
We expect that the original fixed point of $\alpha$ and $\beta$ (Eq.~\eqref{eq:fixed}) will be shifted slightly as follows:
\begin{equation}
(m^\ast_{00}, m^\ast_{01}, m^\ast_{10}, m^\ast_{11}) = (1-\epsilon^\ast_{00}, 1-\epsilon^\ast_{01}, 1-\epsilon^\ast_{10}, 1-\epsilon^\ast_{11}),
\label{eq:shifted}
\end{equation}
where $\left\| \epsilon^\ast_{ij} \right\| \ll 1$, \track{and the asterisk means stationarity. The subscripts $0$ and $1$ mean the mutant and the resident, respectively.
As $m_{ij}$ was defined as player $i$'s assessment of player $j$ in Sect.~\ref{sec:dynamics}, with slight abuse of notation, here we define $m_{00}$ as the assessment between mutant players, $m_{01}$ as a mutant's assessment of a resident player, and so on.}
In short, we have$(\alpha_1, \beta_1) = (\alpha, \beta)$ and
$(\alpha_0, \beta_0) = (\alpha-\delta, \beta-\eta)$ together with Eq.~\eqref{eq:fixed}.
Equation~\eqref{eq:dynamics} is then written as
\begin{equation}
\begin{aligned}
    \frac{{\text{d}}m_{00}}{{\text{d}}t} = qf_1\\
    \frac{{\text{d}}m_{01}}{{\text{d}}t} = qf_2\\
    \frac{{\text{d}}m_{10}}{{\text{d}}t} = qf_3\\
    \frac{{\text{d}}m_{11}}{{\text{d}}t} = qf_4,
\end{aligned}
\label{eq:4dim}
\end{equation}
where
\begin{equation}
    \begin{aligned}
        f_1 \equiv -m_{00}+p\alpha_{0}[m_{00},\beta_{0}(m_{00},m_{00}),m_{00}]+\bar{p}\alpha_{0}[m_{00},\beta_{0}(m_{00},m_{01}),m_{01}] \\
        f_2 \equiv -m_{01}+p\alpha_{0}[m_{01},\beta_{1}(m_{11},m_{10}),m_{00}]+\bar{p}\alpha_{0}[m_{01},\beta_{1}(m_{11},m_{11}),m_{01}] \\
        f_3 \equiv -m_{10}+p\alpha_{1}[m_{10},\beta_{0}(m_{00},m_{00}),m_{10}]+\bar{p}\alpha_{1}[m_{10},\beta_{0}(m_{00},m_{01}),m_{11}] \\
        f_4 \equiv
        -m_{11}+p\alpha_{1}[m_{11},\beta_{1}(m_{11},m_{10}),m_{10}]+\bar{p}\alpha_{1}[m_{11},\beta_{1}(m_{11},m_{11}),m_{11}],
    \end{aligned}
    \label{eq:fi}
\end{equation}
\track{where $\bar{p} \equiv 1-p$ means the fraction of players using the resident norm}. The system is stationary when $f_i=0$.
For example, let us consider IS: If we define a reputational mutant
by $\delta(x,y,z)=\delta_1 xyz$ ($0 < \delta_1 \ll 1$) and explicitly solve the stationarity condition of
Eq.~\eqref{eq:4dim} with a finite value of $p$, the solution is $m^\ast_{ij}=0$ for every pair of $i$ and $j$,
according to a symbolic-algebra package~\cite{mathematica},
although this solution depends on the specific form of $\delta(x,y,z)$.
This result shows that this simple norm does not resist invasion of mutants because of the lack of restoring force against error.
By contrast, SS has a nontrivial fixed point different from
$m^\ast_{ij}=0$ when $\delta(x,y,z)=\delta_1 xyz$, and it is insensitive to the specific form of $\delta(x,y,z)$.

Before proceeding to the next section, let us mention that our perturbative calculation fails when the original fixed point is unstable with respect to assessment error, i.e., when $Q>0$:
It is reasonable to expect that the new fixed point will inherit the stability of the original one, as long as mutants do not create any discernible impact on the system with such a tiny fraction.
Note that we nevertheless keep working with a four-dimensional system including $m_{00}$, the assessment among mutants.
The difference from the original system should be manifested most clearly in this variable, whose initial value is $\alpha_0(1,1,1) = \alpha(1,1,1)-\delta(1,1,1) < 1$.
The mutants will assign worse and worse reputations to each other as time goes by, and the reason is that the mutant norm is almost identical to the resident one, under which any small deviation from perfect reputation tends to grow.
It implies that $\epsilon_{00}^\ast$ may be of $O(1)$ in the end, which undermines the premise of our perturbative approach.
The above argument therefore suggests that we may exclude L2, L5, L6, and L8 from consideration in analyzing the new stationary state that emerges after mutation \track{(see the discussion below Eq.~\eqref{eq:Q}).}
Below, we will see how those four norms actually differ in their behavior from the prediction of our perturbative analysis.

\section{Results}

\subsection{Perturbative expansion}

\subsubsection{Linear order}

The linear-order analysis is straightforward~\cite{lee2021local}:
By expanding Eq.~\eqref{eq:fi} to the first order of $\epsilon_{ij}$'s and equating it with zero, we obtain a linear-algebraic system, which is solved by
\begin{equation}
\begin{aligned}
\epsilon^\ast_{00} &=
\frac{\delta_1 \left\{ A_x^2+A_x (A_y
B_x+A_z-2)-\bar{p}A_y^2 B_x
B_y-\bar{p}A_y^2 B_y^2\right\}}
{(1-A_x-A_z)
(1-A_x-A_y B_x) (1-A_x-A_y B_x-A_y
B_y-A_z)}\\
&+\frac{\delta_1 \left\{ A_z [A_y
(pB_x-\bar{p}B_y)-1]-A_y B_x+1\right\}}
{(1-A_x-A_z)
(1-A_x-A_y B_x) (1-A_x-A_y B_x-A_y
B_y-A_z)}\\
&+\frac{A_y \eta_1
(1-A_x-A_z) (1-A_x-A_y B_x
-\bar{p}A_y B_y-\bar{p}A_z)}
{(1-A_x-A_z)
(1-A_x-A_y B_x) (1-A_x-A_y B_x-A_y
B_y-A_z)}\\
\epsilon^\ast_{01} &=
\frac{A_y \eta_1 p (1-A_x-A_z) (A_y
B_y+A_z)}
{(1-A_x-A_z)
(1-A_x-A_y B_x) (1-A_x-A_y B_x-A_y
B_y-A_z)}\\
&
+\frac{\delta_1 \left[A_x^2+A_x (2 A_y
B_x+A_y B_y+A_z-2)
\right]}
{(1-A_x-A_z)
(1-A_x-A_y B_x) (1-A_x-A_y B_x-A_y
B_y-A_z)}\\
&
+\frac{\delta_1 \left[A_y^2 B_x^2+A_y^2 B_x
B_y p+A_y^2 B_x B_y+A_y^2 B_y^2 p
\right]}
{(1-A_x-A_z)
(1-A_x-A_y B_x) (1-A_x-A_y B_x-A_y
B_y-A_z)}\\
&
+\frac{\delta_1 \left\{
A_z [A_y
(pB_x+B_x+pB_y)-1]-2 A_y B_x-A_y
B_y+1\right\}}
{(1-A_x-A_z)
(1-A_x-A_y B_x) (1-A_x-A_y B_x-A_y
B_y-A_z)}\\
\epsilon^\ast_{10} &=
\frac{A_y (1-A_x-A_y B_x-\bar{p}A_y
B_y-\bar{p}A_z) [\eta_1 (1-A_x-A_z)+(B_x
+B_y) \delta_1]}
{(1-A_x-A_z)
(1-A_x-A_y B_x) (1-A_x-A_y B_x-A_y
B_y-A_z)}\\
\epsilon^\ast_{11} &=
\frac{A_y p (A_y B_y+A_z) (\eta_1
(1-A_x-A_z)+(B_x +B_y) \delta_1)}
{(1-A_x-A_z)
(1-A_x-A_y B_x) (1-A_x-A_y B_x-A_y
B_y-A_z)},
\end{aligned}
\label{eq:shift}
\end{equation}
where
$\delta_1 \equiv \delta(1,1,1)$ and $\eta_1 \equiv \eta(1,1)$.

From Eq.~\eqref{eq:shift}, we can calculate the characteristics of the stationary state. For example, the average cooperation rate is largely determined by $\epsilon_{11}^{\ast}$ when $p$ is small. If we look at the payoff difference between a resident and a mutant, $\Delta \pi_0 \equiv \pi_0 - \pi_1$, it is expressed as
\begin{eqnarray}
    \Delta \pi_0 &=& \left\{ b[p \beta_0 (m_{00}^{\ast}, m_{00}^{\ast})] + (1-p) \beta_1(m_{11}^{\ast}, m_{10}^{\ast})] \right.\nonumber\\
    && \left.- c[p \beta_0(m_{00}^{\ast}, m_{00}^{\ast}) + (1-p) \beta_0(m_{00}^{\ast}, m_{01}^{\ast})] \right\}\nonumber\\
    &-&\left\{ b[p \beta_0 (m_{00}^{\ast}, m_{01}^{\ast})] + (1-p) \beta_1(m_{11}^{\ast}, m_{11}^{\ast})] \right.\nonumber\\
    && \left.- c[p \beta_1(m_{11}^{\ast}, m_{10}^{\ast}) + (1-p) \beta_1(m_{11}^{\ast}, m_{11}^{\ast})] \right\}\\
    &\approx& \left\{ b[p(1-B_x \epsilon_{00}^{\ast}-B_y \epsilon_{00}^{\ast}-\eta_1) + (1-p)(1-B_x \epsilon_{11}^{\ast} - B_y \epsilon_{10}^{\ast})]\right.\nonumber\\
    &&\left. -c [p(1-B_x \epsilon_{00}^{\ast}-B_y \epsilon_{00}^{\ast}-\eta_1) + (1-p)(1-B_x \epsilon_{00}^{\ast} - B_y \epsilon_{01}^{\ast} - \eta_1)]\right\}\nonumber\\
    &-& \left\{ b[p(1-B_x \epsilon_{00}^{\ast}-B_y \epsilon_{01}^{\ast}-\eta_1) + (1-p)(1-B_x \epsilon_{11}^{\ast} - B_y \epsilon_{11}^{\ast})]\right.\nonumber\\
    &&\left. -c [p(1-B_x \epsilon_{11}^{\ast}-B_y \epsilon_{10}^{\ast}) + (1-p)(1-B_x \epsilon_{11}^{\ast} - B_y \epsilon_{11}^{\ast})]\right\}.
    \label{eq:dpi0}
\end{eqnarray}
If Eq.~\eqref{eq:shift} is substituted here,
it turns out that
a mutant becomes worse off than a resident at this stationary state by making $\Delta \pi_0 < 0$,
when cooperation is beneficial enough to satisfy
\begin{equation}
\frac{b}{c} > \frac{1-A_x}{A_y B_y},
\label{eq:bc}
\end{equation}
for any mutant norm that is close to the existing one~\cite{lee2021local}. For the leading eight and IS, this inequality reduces to $b/c > 1$ (Table~\ref{tab:first}), which is exactly the prerequisite for the donation game.

\subsubsection{Second order}

\begin{table}[h]
\begin{center}
\begin{minipage}{\textwidth}
    \caption{Second-order derivatives of $\alpha(x,y,z)$ and $\beta(x,y)$ at $(x,y,z)=(1,1,1)$}
    \label{tab:second}
\begin{tabular}{@{}cccccccccc@{}}
\toprule                                                          
        Norm & $A_{xx}$ & $A_{xy}$ & $A_{zx}$ & $A_{yy}$ & $A_{yz}$ & $A_{zz}$ & $B_{xx}$ & $B_{xy}$ & $B_{yy}$ \\\midrule
        L1 & 0 & 0 &  0 & 0 & 1 & 0 & 0 & 1 & 0\\
        L2 & 0 & 0 &  1 & 0 & 2 & 0 & 0 & 1 & 0\\
        L3 & 0 & 0 &  0 & 0 & 1 & 0 & 0 & 0 & 0\\
        L4 & 0 & 0 & - 1 & 0 & 1 & 0 & 0 & 0 & 0\\
        L5 & 0 & 0 &  1 & 0 & 2 & 0 & 0 & 0 & 0\\
        L6 & 0 & 0 &  0 & 0 & 2 & 0 & 0 & 0 & 0\\
        L7 & 0 & 0 & - 1 & 0 & 1 & 0 & 0 & 0 & 0\\
        L8 & 0 & 0 &  0 & 0 & 2 & 0 & 0 & 0 & 0\\
        IS & 0 & 0 &  0 & 0 & 0 & 0 & 0 & 0 & 0\\
\botrule
\end{tabular}
\end{minipage}
\end{center}
\end{table}

Although the solution in Eq.~\eqref{eq:shift} explains the response of a variety of norms,
we have at least two reasons to conduct a higher-order analysis: First, Eq.~\eqref{eq:shift} actually fails to locate the nontrivial fixed point when applied to L1, L3, L4, L7, and IS because they have $Q=0$, making the denominators of Eq.~\eqref{eq:shift} vanish.
Second, the difference between SS (Eq.~\eqref{eq:ss0}) and IS (Eq.~\eqref{eq:scoring}) first arises in
their second-order derivatives in the sense that $A_{yz}=1$ for SS whereas
$A_{yz}=0$ for IS,
where $A_{\mu \nu} \equiv \left. \partial^2 \alpha / \partial\mu \partial\nu
\right\vert_{(1,1,1)}$ and $B_{\mu\nu} \equiv \left.
\partial^2 \beta /\partial \mu \partial \nu \right\vert_{(1,1)}$ (Table~\ref{tab:second}). We may thus say that $A_{yz}$ is the parameter related to justified punishment, the key difference between SS and IS.

Throughout this work, we will assume that $\delta$, $\eta$, and
$\epsilon_{ij}$'s as well as their partial derivatives are all
small parameters of the same order of magnitude.
If we take into account second-order terms so as to obtain sensible estimates of $\epsilon_{ij}$'s (Appendix~\ref{app:second}),
we have to solve a set of four coupled nonlinear equations. We use
Newton's method to tackle this problem (Appendix~\ref{app:newton}),
and the trial solution
$(\hat{\varepsilon_{00}},\hat{\varepsilon_{01}},\hat{\varepsilon_{10}},\hat{\varepsilon_{11}})$
is chosen from the linear solution (Eq.~\eqref{eq:shift}) in the limit of $p\to 0$:
\begin{equation}
\begin{aligned}
\hat{\epsilon}_{00} &= \frac{(1-A_x+A_y B_y)\delta_1 +
(1-A_x-A_z)A_y \eta_1}{(1-A_x-A_z)(1-A_x-A_y B_x)}\\
\hat{\epsilon}_{01} &= \frac{\delta_1}{1-A_x-A_z}\\
\hat{\epsilon}_{10} &= \frac{(B_x+B_y)\delta_1 +
(1-A_x-A_z)\eta_1}{(1-A_x-A_z)(1-A_x-A_y B_x)} A_y\\
\hat{\epsilon}_{11} &= 0.
\end{aligned}
\label{eq:epsilons}
\end{equation}
This trial solution is free from the problem of vanishing denominators for L1, L3, L4, L7, and IS, and the $p$-dependency will be recovered later by applying Newton's method.
Still, Eq.~\eqref{eq:epsilons} fails to distinguish SS from IS because it is written only in terms of first-order derivatives.
We additionally introduce a small regularization parameter $\omega$ so that
\begin{equation}
(A_x, A_y, A_z, B_x, B_y) = (0, 1-\omega, 0, 0, 1-\omega),
\label{eq:omega}
\end{equation}
which reproduces Eq.~\eqref{eq:pure} in the limit of $\omega\to 0$.
This parameter can be interpreted as generosity~\cite{park2017role,okada2020two}: \track{for example, this regularization changes $\alpha(1,1,0)$ from zero to $\omega>0$, meaning that a donation to an ill-reputed recipient is not judged as completely bad. Mathematically, this parameter helps to understand the divergence in Eq.~\eqref{eq:shift} in the following way}: when substituted in Eq.~\eqref{eq:shift}, Eq.~\eqref{eq:omega} implies
\begin{align}
    \begin{pmatrix}
    \varepsilon_{00}^\ast \\
    \varepsilon_{01}^\ast \\
    \varepsilon_{10}^\ast \\
    \varepsilon_{11}^\ast
    \end{pmatrix}
    =
    \frac{1}{2\omega/\delta_1}
    \begin{pmatrix}
    p+4\omega \\
    p+2\omega \\
    p+2\omega \\
    p
    \end{pmatrix},
    \label{eq:approx1}
\end{align}
which diverges as $\omega \to 0$. Note that we have chosen $\eta=0$ to focus on mutation in $\alpha$.

To incorporate second-order effects such as justified punishment, we now express Eq.~(\ref{eq:fi}) as second-order polynomials in $\epsilon_{ij}$'s (Appendix~\ref{app:second}) and plug them into Newton's method (Appendix~\ref{app:newton}) as follows:
\begin{align}
    \begin{pmatrix}
    \varepsilon_{00}^\ast \\
    \varepsilon_{01}^\ast \\
    \varepsilon_{10}^\ast \\
    \varepsilon_{11}^\ast
    \end{pmatrix}
    \approx
    \begin{pmatrix}
    \hat{\varepsilon}_{00} \\
    \hat{\varepsilon}_{01} \\
    \hat{\varepsilon}_{10} \\
    \hat{\varepsilon}_{11}
    \end{pmatrix}
    -
    \begin{pmatrix}
    \frac{\partial f_{1}}{\partial \varepsilon_{00}}&\cdots&\frac{\partial f_{1}}{\partial \varepsilon_{11}} \\ \vdots&\ddots&\vdots \\
    \frac{\partial f_{4}}{\partial \varepsilon_{00}}&\cdots&\frac{\partial f_{4}}{\partial \varepsilon_{11}}\end{pmatrix}^{-1}
    \begin{pmatrix}
    f_{1}(\hat{\varepsilon}_{00},\hat{\varepsilon}_{01},\hat{\varepsilon}_{10},\hat{\varepsilon}_{11}) \\
    f_{2}(\hat{\varepsilon}_{00},\hat{\varepsilon}_{01},\hat{\varepsilon}_{10},\hat{\varepsilon}_{11}) \\
    f_{3}(\hat{\varepsilon}_{00},\hat{\varepsilon}_{01},\hat{\varepsilon}_{10},\hat{\varepsilon}_{11}) \\
    f_{4}(\hat{\varepsilon}_{00},\hat{\varepsilon}_{01},\hat{\varepsilon}_{10},\hat{\varepsilon}_{11})
    \end{pmatrix},
    \label{eq:newton}
\end{align}
where the inverse matrix is evaluated at the trial solution in Eq.~\eqref{eq:epsilons}.
After discarding high-order terms of $\omega$, $p$, and $\delta_1$ in the numerators and denominators, we obtain
\begin{align}
    \begin{pmatrix}
    \varepsilon_{00}^\ast \\
    \varepsilon_{01}^\ast \\
    \varepsilon_{10}^\ast \\
    \varepsilon_{11}^\ast
    \end{pmatrix}
    \approx
    \frac{1}{K+2\omega/\delta_1}
    \begin{pmatrix}
    p+4\omega \\
    p+2\omega \\
    p+2\omega \\
    p
    \end{pmatrix}.
    \label{eq:approx2}
\end{align}
where
\begin{equation}
    K \equiv [3+4(A_{yz}+A_{zx}+B_{xy})]p+(6A_{zx}+2B_{xy}+6)\omega - \omega^2/\delta_1.
    \label{eq:K}
\end{equation}
In deriving this expression, we have used $A_{xx} = A_{yy} = A_{zz} = 0$ (Table~\ref{tab:second}). The comparison between Eqs.~\eqref{eq:approx1} and \eqref{eq:approx2} shows how
the denominators are modified by incorporating second-order effects.

\track{
However, the stability threshold of $b/c$ (Eq.~\eqref{eq:bc})
will be insensitive to this modification because all of $\epsilon_{ij}^{\ast}$'s
change by exactly the same factor at this level of approximation.
It implies that the resident norms in Table~\ref{tab:cont} under consideration will suppress the invasion of close mutants as long as the donation game constitutes the PD by having $b>c>0$ in the limit of $\omega \to 0$.
Of course, we have to stress that Eq.~\eqref{eq:approx2} only provides a rough estimate,
and the performance of this approximation will be checked with MC calculations below.}
In particular, Eq.~\eqref{eq:approx2} does not reproduce Eq.~\eqref{eq:approx1} even if we set all the second derivatives to zero, and the remaining terms in the denominator seem to be an artifact of Newton's method.

\subsection{Numerical calculation}\label{sec2}

We have carried out MC calculations to check the validity of Eq.~\eqref{eq:approx2}. We consider a population of size $N=50$. The system shows little size dependence in a stationary state~\cite{lee2021local}.
The resident norm is basically obtained from the tri- and bi-linear interpolations of Table~\ref{tab:eight}, but we have to include the regularization parameter $\omega$ [see Eq.~\eqref{eq:omega}], which means that both $\alpha_{1D1}$ and $\beta_{10}$ have to be identified with $\omega$ instead of zero. The results are $\alpha_{\text{SS}} = yz-z+1+\omega zx(1-y)$, $\alpha_{\text{IS}} = y+\omega zx(1-y)$, \track{and} $\beta_{\text{SS}} = \beta_{\text{IS}} = y + \omega x(1-y)$.
A mutant norm is defined by $\delta(x,y,z)$, the difference from the resident norm, and we choose it to be $\delta(x,y,z) = \delta_1 xyz$ with $\delta_1 = 2\times 10^{-2}$, \track{whereas $\eta \equiv 0$ because we wish to focus on reputational mutants.}
The number of mutants ranges from one to five, i.e., from $p=0.02$ to $0.1$, but we will show only the results for $p=0.02$ below because the number of mutants makes no qualitative difference.
Every player $i$ in the population has her own norm $(\alpha_i,\beta_i)$, which is either $(\alpha,\beta)$ or $(\alpha_0, \beta_0)=(\alpha-\delta, \beta)$, and we have an $N \times N$ image matrix $\{ m_{ij} \}$. The simulation procedure goes as follows:
\begin{enumerate}
    \item Initialize the image matrix by setting every element to one. The time index $t$ is set to zero.
    \item Pick up a pair of players randomly from the population, one as a donor and the other as a recipient. Let the donor and the recipient be denoted as $i$ and $j$, respectively.
    \item The donor chooses how much to cooperate to the recipient by calculating $\beta_i (m_{ii}, m_{ij})$. This interaction is observed by themselves with certainty, and by other players with probability $q$. As argued above, we set $q=1$ for the sake of simplicity.
    \begin{itemize}
    \item If player $k$ observes the interaction, update the matrix element $m_{ki}$ to $\alpha_k \left[ m_{ki}, \beta_i(m_{ii}, m_{ij}), m_{kj} \right]$.
    \item  Otherwise, keep $m_{ki}$ unchanged.
    \end{itemize}
    \item Increase $t$ by one.
    \item If $t$ equals a predefined value $M$, stop this simulation. Otherwise, go back to Step 2. We have chosen $M = 5\times 10^4$, which makes the number of updates per player $M/N = 10^3$.
\end{enumerate}

\begin{figure}[htb]
  \centering
    \includegraphics[width=0.44\textwidth]{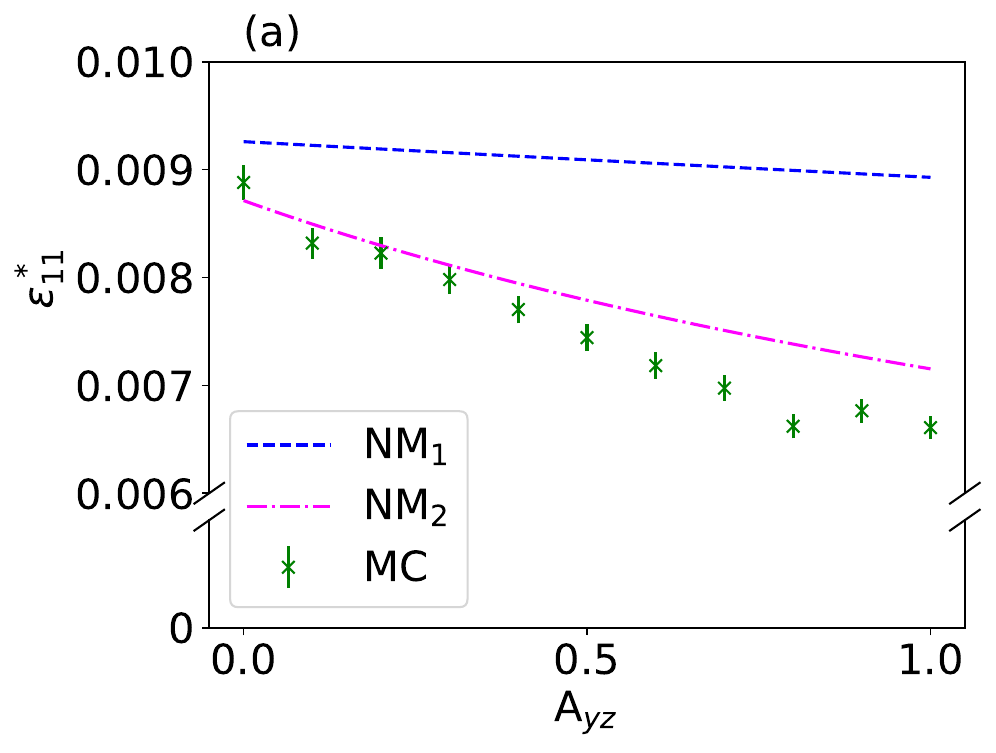}
    \includegraphics[width=0.44\textwidth]{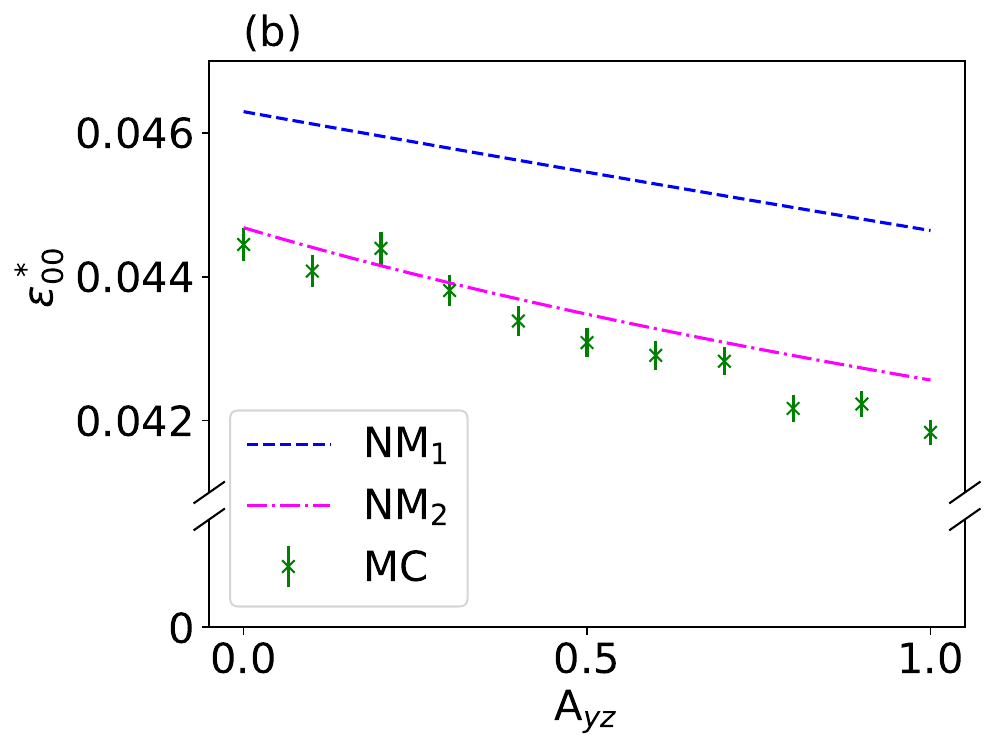}
    \caption{Deviation from the initial state for norms between IS and SS, where IS and SS correspond to $A_{yz}=0$ and $1$, respectively. \track{We have plotted (a) $\epsilon_{11}^{\ast}$ and (b) $\epsilon_{00}^{\ast}$ here and observed similar trends for $\epsilon_{10}^{\ast}$ and $\epsilon_{01}^{\ast}$ as well. By `NM$_1$', we mean Eq.~\eqref{eq:approx2} resulting from Newton's method, which explains the overall sizes of $\epsilon_{ij}^{\ast}$'s and their inverse relationship with $A_{yz}$. The approximation is improved by iterating Newton's method once more, and the results are denoted by NM$_2$.
    In the MC calculation, every data point is an average over $10^3$ samples.
    The regularization parameter has been set to $\omega=0.02$.}}
    \label{fig:newton}
\end{figure}

Figure~\ref{fig:newton} compares the MC results with Newton's method. We are considering norms between IS and SS, obtained by interpolating them as follows:
\begin{subequations}
\begin{align}
\alpha &= (1-u) \alpha_{\text{IS}} + u \alpha_{\text{SS}}\\
\beta &= (1-u) \beta_{\text{IS}} + u \beta_{\text{SS}}
\end{align}
\label{eq:intermediate}
\end{subequations}
where $0\le u \le 1$. Thereby the effect of $A_{yz}$ can be controlled because $A_{yz}$ is identified with $u$ in this parametrization.
As we see in Fig.~\ref{fig:newton}, Newton's method correctly predicts the overall size of $\epsilon_{00}^{\ast}$ as well as its inverse relationship with $A_{yz}$.
The performance is improved further by applying Newton's method iteratively, but with no essential changes in the qualitative behavior.
The iteration becomes more and more important as $\omega \to 0$, where the simple linear-order analysis that we have started with leads to a pathological result.

\begin{figure}[htb]
   \centering
    \includegraphics[width=0.75\textwidth]{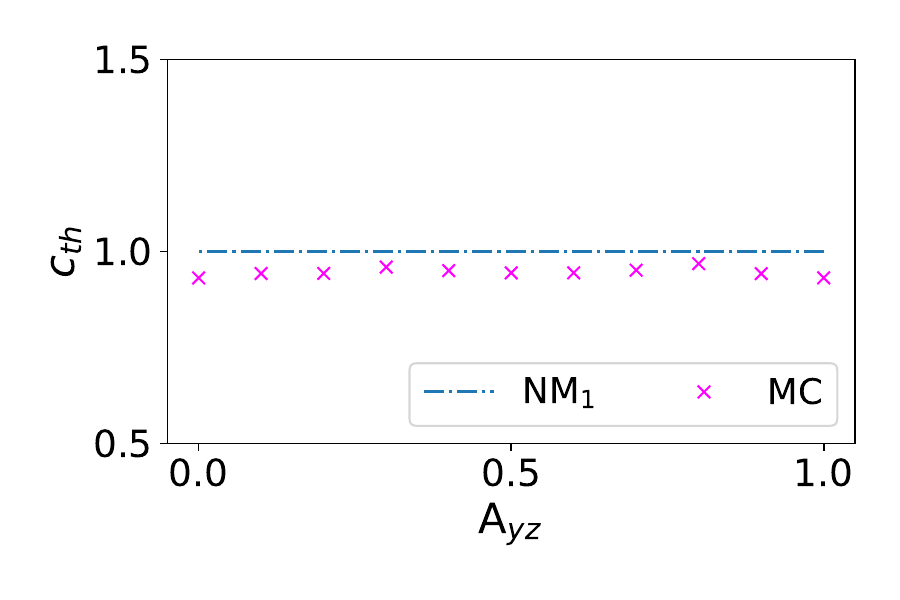}
    \caption{\track{Threshold values of $c$, below which a reputational mutant with $\left\|\delta\right\| \ll 1$ becomes worse off than a resident player in a stationary state. The line denoted by `NM$_1$' shows the result from the direct application of Newton's method (Eq.~\eqref{eq:approx2}), which is $c_\text{th}=1$. Our MC data, the same ones as in Fig.~\ref{fig:newton}, show that this prediction is approximately correct.}
    }
    \label{fig:threshold}
\end{figure}

\track{An important question would be how the stability threshold (Eq.~\eqref{eq:bc}) changes when we take the second-order effects into consideration.
Let us set $b\equiv 1$ as the unit of payoffs.
We define $c_{\text{th}}$ as the threshold of $c$ below which a mutant norm defined by $\left\| \delta \right\| \ll 1$ becomes worse off than a resident player, i.e., by making $\Delta \pi_0 <0$ (Eq.~\eqref{eq:dpi0}).
Equation~\eqref{eq:approx2}, resulting from Newton's method, predicts $c_{\text{th}}=1$, which means that the mutant will always be worse off than a resident player as long as the donation game constitutes the PD by having $c<b\equiv 1$. Note that this prediction coincides with the linear-order prediction (Eq.~\eqref{eq:bc}) in the limit of $\omega\to 0$. Although our approximation agrees with the MC data only qualitatively in terms of $\epsilon_{ij}$'s (see `NM$_1$' in Fig.~\ref{fig:newton}), the threshold estimation shows a better agreement: In Fig.~\ref{fig:threshold}, we have depicted the values of $c_{\text{th}}$ that are calculated from the MC data in Fig.~\ref{fig:newton}. The threshold thereby obtained is actually close to one regardless of $A_{yz}$, which means that mutants will be successfully suppressed by the norm between IS and SS (Eq.~\eqref{eq:intermediate}) unless the dilemma strength becomes extremely high, i.e., $c \approx 1$.
}

\begin{figure}[htb]
   \centering
    \includegraphics[width=0.75\textwidth]{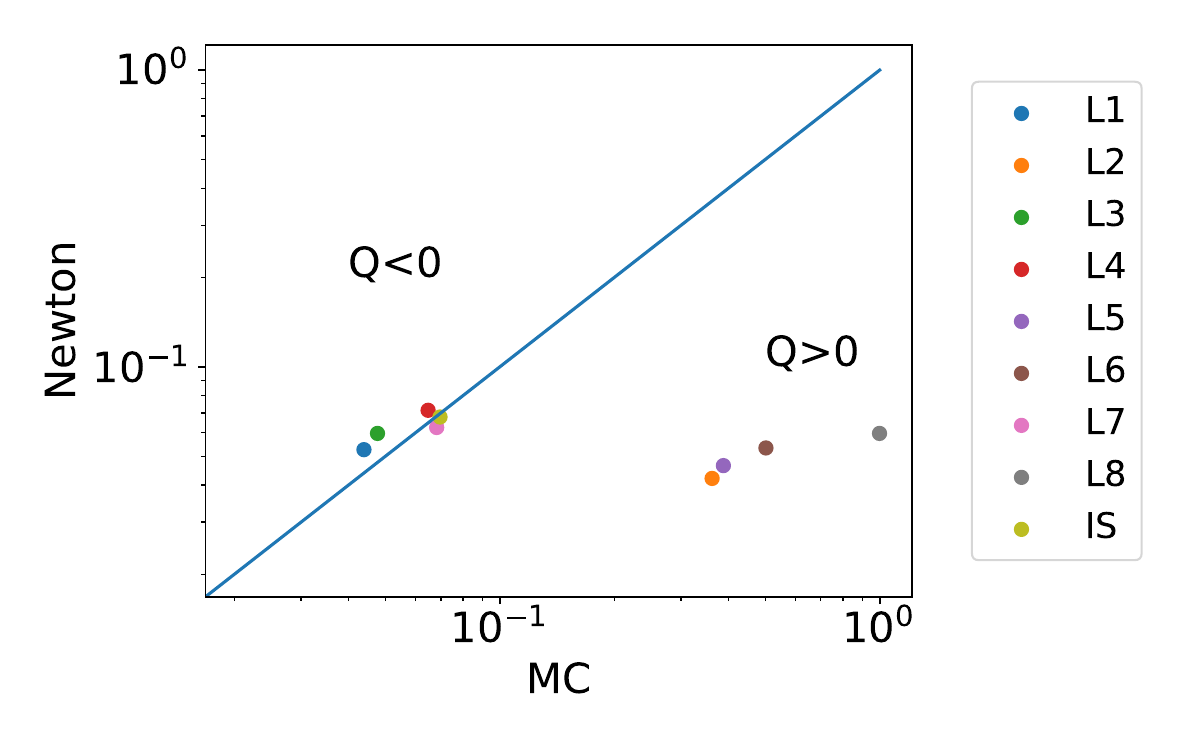}
    \caption{Comparison \track{of $\epsilon_{00}^\ast$} between Newton's method (Eq.~\eqref{eq:approx2}) and MC data for the leading eight and IS, when $\omega=5\times 10^{-3}$. 
    The solid line represents $y=x$ in the $xy$ plane,
    and the four points far off the line denote L2, L5, L6, and L8, respectively. This figure is plotted on a log-log scale, and each data point is obtained from $10^3$ samples.}
    \label{fig:agreement}
\end{figure}

If we include norms other than IS and SS, they have various combinations of second-order derivatives (Table~\ref{tab:second}). By comparing Eq.~\eqref{eq:approx2} with their MC data (Fig.~\ref{fig:agreement}), we see the following points: First, some norms exhibit completely different behavior from the prediction of Eq.~\eqref{eq:approx2}. They are the norms with $Q>0$, i.e., L2, L5, L6, and L8. This result clearly supports our argument in Sect.~\ref{subsec:mutation} that our perturbative approach does not apply to them. Second, Eq.~\eqref{eq:approx2} nevertheless explains the MC data for the other five norms, i.e., L1, L3, L4, L7, and IS, in the presence of such a small regularization parameter.

\section{Summary and discussion}\label{sec12}

In summary, we have extended the perturbative analysis of mutation occurring in the assessment rule to the second order, \track{within the continuum framework for reputation and cooperation that our previous works have developed~\cite{lee2021local,lee2022second}.} We have focused on the leading eight, for which the linear-order theory encounters singularity. \track{The main contribution of this work is that we have demonstrated} how the second-order derivatives regulate the divergence of $\epsilon_{ij}^{\ast}$, in comparison with the linear-order analysis (Eq.~\eqref{eq:approx1}). Our result thereby allows us to identify which characteristics of a resident norm affect the deviation from the initial cooperative state when mutation has occurred.

In our formulation, a norm is characterized by its first- and second-order derivatives, and each derivative can be interpreted as a sensitivity measure~\cite{lee2021local}: The difference between SS and IS lies in $A_{yz} \equiv \left. \partial^2 \alpha(x,y,z) / \partial y \partial z \right\vert_{(1,1,1)}$, which means the response in $\alpha$ when the donor's behavior ($y$) and the recipient's reputation ($z$) change from the reference initial state $(x,y,z)=(1,1,1)$. In short, it represents whether a donor's punishment against her ill-reputed recipient is justified by the resident assessment rule $\alpha$. Likewise, $A_{zx}$ in Eq.~\eqref{eq:K} concerns the response in $\alpha$ when the donor's reputation ($x$) and the recipient's reputation ($z$) change. If an ill-reputed player's cooperation to another ill-reputed player is regarded as bad by $\alpha$, we have $A_{zx}<0$, which decreases the modifier $K$ in the denominators (Eq.~\eqref{eq:K}). It is straightforward to see that $B_{xy}$ is also related to whether an ill-reputed donor has to cooperate to her ill-reputed recipient.

Recall that
the competition between $A_{yz}$ and $A_{zx}$ as well as the stabilizing role of $B_{xy}$ in Eq.~\eqref{eq:K} has already been observed
in the analysis of assessment error~\cite{lee2022second}: Specifically, if we consider the average deviation from the fully cooperative initial state, which may be identified with $\epsilon$ in Sect.~\ref{sec:error}, its time evolution is described by the following dynamics:
\begin{equation}
    \frac{{\text{d}}\epsilon}{{\text{d}}\tau} \approx A_z \epsilon - (A_{yz} + A_{zx} + B_{xy}) \epsilon^2,
\end{equation}
where $\tau \equiv qt$ is a rescaled time variable that has absorbed the observation probability.
The factor inside the parentheses, $A_{yz} + A_{zx} + B_{xy}$, is the same as in the modifier $K$ (Eq.~\eqref{eq:K}). We may furthermore point out that the quantity $Q$ defined in Eq.~\eqref{eq:Q}, originating from the analysis of mutation [see the denominators of Eq.~\eqref{eq:shift}], also characterizes the stability against assessment error in the linear-order analysis~\cite{lee2021local}. These findings again indicate a close relationship between the response to mutation and the stability against assessment error.
The reason can be argued as follows: We can imagine a mutant pretending that the deviation occurs by error. If the initial state is fragile against assessment error, so is it against such a mutant.

For the perturbative analysis to make sense in the first place,
we have required $A_z=0$ to make $Q \le 0$, which means that a well-reputed donor's cooperation toward an ill-reputed player has to be regarded as good. In the second-order analysis, we observe the role of justified punishment, which is missing in IS: Its importance has been pointed out repeatedly in the literature, but we have also found that the effect will be weakened if cooperation between two ill-reputed players is regarded as bad ($A_{zx}<0$) and strengthened if such behavior is encouraged ($B_{xy}>0$). The continuous model of indirect reciprocity thereby provides us with useful insights to systematically understand how social norms should be organized to secure cooperation.

\backmatter

\bmhead{Acknowledgments}
We gratefully acknowledge discussions with Yohsuke Murase.
We appreciate the APCTP for its hospitality during the completion of this work.

\section*{Author contributions}
Formal Analysis, Software, Visualization: YM; Conceptualization, Methodology, Writing – original draft: SKB. 

\subsection*{Funding}
S.K.B. acknowledges support by Basic Science Research Program through the National Research Foundation of Korea (NRF) funded by the Ministry of Education (NRF-2020R1I1A2071670).

\subsection*{Code availability} The source code for this study is available at \url{https://github.com/angcaljine/game_continous_model}.

\subsection*{Declarations}

\subsection*{Conflict of interest}
The authors have no conflict of interest.

\begin{appendices}

\section{Second-order perturbations}\label{app:second}
\setcounter{equation}{0}

The second-order perturbation for $\beta_1$ is written as follows:
\begin{align}
&\beta_1 (m_{11}, m_{1j}) = \beta(1-\epsilon_{11}, 1-\epsilon_{1j})\\
&\approx 1 - B_x \epsilon_{11} - B_y \epsilon_{1j} +
\frac{1}{2} B_{xx} \epsilon_{11}^2 + B_{xy} \epsilon_{11} \epsilon_{1j}
+ \frac{1}{2} B_{yy} \epsilon_{1j}^2 \\
&\equiv 1 - \kappa_1.
\end{align}
Here, we write $\kappa_1 \equiv \kappa_1^{(1)} + \kappa_1^{(2)}$ by defining
\begin{align}
\kappa_1^{(1)} &\equiv B_x \epsilon_{11} + B_y
\epsilon_{1j}\\
\kappa_1^{(2)} &\equiv -
\left( \frac{1}{2} B_{xx} \epsilon_{11}^2 + B_{xy} \epsilon_{11}
\epsilon_{1j} + \frac{1}{2} B_{yy} \epsilon_{1j}^2 \right)
\end{align}
as first- and second-order corrections, respectively.
Similarly,
\begin{align}
&\beta_0 (m_{00}, m_{0j}) = \beta(m_{00}, m_{0j}) - \eta(m_{00}, m_{0j})\\
&= \beta(1-\epsilon_{00}, 1-\epsilon_{0j}) - \eta(1-\epsilon_{00},
1-\epsilon_{0j})\\
&\approx \left(1 - B_x \epsilon_{00} - B_y \epsilon_{0j} +
\frac{1}{2} B_{xx} \epsilon_{00}^2 + B_{xy} \epsilon_{00}
\epsilon_{0j} + \frac{1}{2} B_{yy} \epsilon_{0j}^2 \right)\nonumber\\
&- \left(\eta_1 - \eta_x \epsilon_{00} - \eta_y \epsilon_{0j} \right)\\
&\equiv 1 - \kappa_0,
\end{align}
where $\kappa_0 \equiv \kappa_0^{(1)} + \kappa_0^{(2)}$ with
\begin{subequations}
\begin{align}
\kappa_0^{(1)}
&\equiv B_x \epsilon_{00} + B_y \epsilon_{0j} + \eta_1\\
\kappa_0^{(2)} &\equiv -\left( \frac{1}{2} B_{xx} \epsilon_{00}^2
+ B_{xy} \epsilon_{00} \epsilon_{0j} + \frac{1}{2} B_{yy}
\epsilon_{0j}^2 \right) - (\eta_x \epsilon_{00} + \eta_y \epsilon_{0j}).
\end{align}
\end{subequations}
The second-order perturbation for $\alpha_1$ is also straightforward:
\begin{align}
&\alpha_1 [m_{1i}, \beta_i (m_{ii}, m_{ij}), m_{1j}]
\approx \alpha(1-\epsilon_{1i}, 1-\kappa_i, 1-\epsilon_{1j})\\
&\approx 1 - A_x \epsilon_{1i} - A_y \kappa_i - A_z
\epsilon_{1j}
+ \frac{1}{2} A_{xx} \epsilon_{1i}^2 + \frac{1}{2} A_{yy}
\left(\kappa_i^{(1)} \right)^2 + \frac{1}{2} A_{zz}
\epsilon_{1j}^2\nonumber\\
&+ A_{xy} \epsilon_{1i} \kappa_i^{(1)}
+ A_{yz} \kappa_i^{(1)} \epsilon_{1j}
+ A_{zx} \epsilon_{1i} \epsilon_{1j}.
\end{align}
Likewise, we have
\begin{align}
&\alpha_0 [m_{0i}, \beta_i (m_{ii}, m_{ij}), m_{0j}]\\
&\approx \alpha(1-\epsilon_{0i}, 1-\kappa_i, 1-\epsilon_{0j}) -
\delta(1-\epsilon_{0i}, 1-\kappa_i, 1-\epsilon_{0j})\\
&\approx \left[ 1 - A_x \epsilon_{0i} - A_y \kappa_i - A_z
\epsilon_{0j}
+ \frac{1}{2} A_{xx} \epsilon_{0i}^2 + \frac{1}{2} A_{yy}
\left(\kappa_i^{(1)} \right)^2 + \frac{1}{2} A_{zz}
\epsilon_{0j}^2 \right.\nonumber\\
& \left. + A_{xy} \epsilon_{0i} \kappa_i^{(1)}
+ A_{yz} \kappa_i^{(1)} \epsilon_{0j}
+ A_{zx} \epsilon_{0i} \epsilon_{0j}
\nonumber \right]\\
&- \left( \delta_1 - \delta_x \epsilon_{0i} - \delta_y \kappa_i^{(1)} -
\delta_z \epsilon_{0j} \right).
\end{align}

\section{Newton's method}\label{app:newton}
\setcounter{equation}{0}

Assume that we have to solve an equation
\begin{equation}
0 = f (x)
\end{equation}
for which we have an approximate solution $\hat{x}$.
By expanding the above equation around $\hat{x}$,
we see
\begin{equation}
0 = f(x^\ast) = f (\hat{x}) + (x^\ast - \hat{x})
\left.\frac{{\text{d}}f}{{\text{d}}x}\right\vert_{\hat{x}}
+ \frac{1}{2} (x^\ast-\hat{x})^2 \left. \frac{{\text{d}}^2f}{{\text{d}}x^2}\right\vert_{\hat{x}} + \ldots,
\end{equation}
from which we obtain
\begin{equation}
x^\ast = \left[\hat{x} - \frac{f(\hat{x})}{\left. {\text{d}}f/{\text{d}}x \right\vert_{\hat{x}}}
\right] - \frac{1}{2} (x^\ast-\hat{x})^2
\frac{\left. {\text{d}}^2f/{\text{d}}x^2 \right\vert_{\hat{x}}}{\left. {\text{d}}f/{\text{d}}x \right\vert_{\hat{x}}} + \ldots.
\end{equation}
As a result, the improved approximation, the term inside the square brackets on
the right-hand side, deviates from the actual solution by
\begin{equation}
x^\ast - \left[\hat{x} - \frac{f(\hat{x})}{\left.{\text{d}}f/{\text{d}}x \right\vert_{\hat{x}}}
\right] \approx \left( -
\frac{\left. {\text{d}}^2f/{\text{d}}x^2 \right\vert_{\hat{x}}}{2\left. {\text{d}}f/{\text{d}}x \right\vert_{\hat{x}}}
\right)
(x^\ast-\hat{x})^2.
\end{equation}
The size of error in this improved estimate is thus proportional to the square
of the previous error. Setting the right-hand side to zero leads to Newton's method to obtain an improved estimate $x^\ast$ from the trial solution $\hat{x}$ in a one-dimensional problem. It is straightforward to generalize this formula to a multi-dimensional case and derive Eq.~\eqref{eq:newton} in the main text.

\section{\track{Bi- and tri-linear interpolation}}\label{app:interpolation}
\setcounter{equation}{0}

\track{The bi-linear interpolation of $\beta$ assumes the following functional form:
\begin{equation}
    \beta(x,y) = a_0 + a_1 x + a_2 y + a_3 xy,
\end{equation}
where $a_i$'s are constant coefficients to be found. If we take L1 as an example, we know from Table~\ref{tab:eight} that
\begin{subequations}
\begin{align}
    1 &= \beta_{00} = \beta(0,0) = a_0\\
    0 &= \beta_{10} = \beta(1,0) = a_0+a_1\\
    1 &= \beta_{01} = \beta(0,1) = a_0+a_2\\
    1 &= \beta_{11} = \beta(1,1) = a_0+a_1+a_2+a_3,
\end{align}
\label{eq:bilinear}
\end{subequations}
where $C$ and $D$ are identified with $1$ and $0$, respectively. Solving Eq.~\eqref{eq:bilinear}, we get $(a_0, a_1, a_2, a_3) = (1,-1,0,1)$, or, equivalently, $\beta(x,y) = 1 - x + xy$, as can be found in Table~\ref{tab:cont}.
Likewise, the tri-linear interpolation begins by assuming
\begin{equation}
    \alpha(x,y,z) = a_0 + a_1x + a_2 y + a_3 z + a_4 xy + a_5 yz + a_6 zx + a_7 xyz.
\end{equation}
By fixing the eight endpoints as specified in Table~\ref{tab:eight}, we obtain a set of eight coupled linear equations of $a_0$, $a_1$, $\ldots$, and $a_7$, which can readily be solved to determine $\alpha(x,y,z)$.
}

\end{appendices}

\nolinenumbers

\bibliography{sn-bibliography}

\end{document}